\documentstyle[12pt,titlepage]{article}

\font\grande=cmr10 scaled \magstep4
\font\medio=cmr10 scaled \magstep2
\outer\def\beginsection#1\par{\medbreak\bigskip
      \message{#1}\leftline{\bf#1}\nobreak\medskip
\vskip-\parskip
      \noindent}

\def\laq{\raise 0.4ex\hbox{$<$}\kern -0.8em\lower 0.62
ex\hbox{$\sim$}}
\def\gaq{\raise 0.4ex\hbox{$>$}\kern -0.7em\lower 0.62
ex\hbox{$\sim$}}
\def\beq{\begin{equation}}
\def\eeq{\end{equation}}
\def\bea{\begin{eqnarray}}
\def\eea{\end{eqnarray}}
\def\bean{\begin{eqnarray*}}
\def\eean{\end{eqnarray*}}

\def \bk {{\bf k}}

\def \Ga {\Gamma}

\def \sg {\sigma}
\def \da {\delta}

\def \r {\rho}

\def \Om {\Omega}

\def \al {\alpha}

\def \Ga {\Gamma}

\def \sg {\sigma}
\def \Sg {\Sigma}
\def \si {\sigma}
\def \da {\delta}

\def \r {\rho}

\def \Om {\Omega}

\begin{document}
\bibliographystyle {unsrt}

\titlepage
\begin{flushright}
DFTT-27/98 \\
CERN-TH/98-169 \\
astro-ph/9806015 \\

\end{flushright}
\vspace{15mm}
\begin{center}
{\grande Massless (pseudo-)scalar seeds of CMB anisotropy}\\

\vspace{15mm}

 R. Durrer${}^{(1)}$, M. Gasperini${}^{(2)}$,
  M. Sakellariadou${}^{(1),(3)}, $ G. Veneziano${}^{(4)}$ \\
\vspace{6mm}
${}^{(1)}$
{\sl  D\'epartement de Physique Th\'eorique, Universit\'e de
Gen\`eve, \\
24 quai E. Ansermet,  CH-1211 Geneva, Switzerland }\\
${}^{(2)}$
{\sl Dipartimento di Fisica Teorica, Universit\`a di Torino, \\
Via P. Giuria 1, 10125 Turin, Italy }\\
${}^{(3)}$
{\sl D\'epartement d'Astrophysique Relativiste et de Cosmologie, \\
UPR 176 du Centre National de la Recherche Scientifique, \\
Observatoire de Paris, 92195 Meudon, France }\\
${}^{(4)}$
{\sl Theory Division, CERN, CH-1211 Geneva 23, Switzerland} \\
\end{center}

\vskip 2cm
\centerline{\medio  Abstract}

\noindent
A  primordial stochastic background of very weakly coupled
massless (pseudo-)scalars can seed CMB
anisotropy, when  large-scale fluctuations of
their stress-tensor re-enter the
horizon during the matter-dominated era.
A general relation between  multipole
coefficients of the CMB anisotropy and the seed's energy spectrum is  
derived.
Magnitude and tilt of the observed  anisotropies
can be reproduced for the nearly scale-invariant axion spectra that are
predicted in a particularly symmetric class of string cosmology
backgrounds.
\vspace{5mm}

\vfill
\begin{flushleft}
CERN-TH/98-169\\
May 1998\\

\end{flushleft}

\newpage

In this letter we point out a possible new mechanism
for generating  large-scale CMB anisotropies.
We will show that
a cosmologically amplified stochastic background of
massless (pseudo-)scalar perturbations,
if primordially
produced with a properly normalized
 nearly scale-invariant spectrum, can seed CMB
 temperature anisotropies in a way consistent with  observations
\cite{COBE}. Axionic perturbations with the needed
characteristics can be produced \cite {Copeland}, \cite{Buon}, \cite{Hadad},
for instance,  in
the so-called pre-big bang scenario  \cite{PBB} of string
cosmology.

For the sake of generality we define a massless (pseudo-)scalar
``seed''  field $\si$ through the way it enters the  low-energy
effective action:
\beq
S_{eff} = - {1\over 2} \int d^4 x ~ \sqrt{-g}~ A~ (\partial_\mu
\si)^2 , \label{action}
\eeq
and by the two additional conditions:
\beq
 \langle\si\rangle = 0~~~~~ , ~~~~~~~~
~~~~~~\Omega_{\si} \ll 1.
\label{def}
\eeq
In Eq.~(\ref{action}),  $A$ is, in general, a $\si$-independent
scalar combination of fields  providing, together with the
metric, the cosmological background.
In Eq.~(\ref{def}), the brackets denote spatial average (or
expectation value if perturbations are quantized), and
$\Omega_{\si}$ is the fraction  of critical energy density
carried by the seed field. Such a fraction being small,
 seeds do not  influence the
 background itself. The above conditions, together with the
restriction to massless seeds, make it essentially  mandatory for
such   seeds, if they exist, to
consist of very weakly coupled pseudo-scalar (rather than
scalar)   particles. An
example would be the gravitationally coupled  ``universal axion''
of string theory, on which we shall come back at the end of this letter.

 We have in mind, typically, a situation  in which, in  the absence of
$\si$, large-scale CMB anisotropies directly induced by quantum
fluctuations  of the metric are too small; we want to investigate
under which conditions  seeds may  provide the  dominant source for
them. Seed vacuum
fluctuations, after being amplified outside the horizon during
inflation,  re-enter during the standard Friedman-Robertson-Walker  
(FRW) era as
stochastic Gaussian fields, and give rise to  non-trivial  -- and
not necessarily
Gaussian -- fluctuations of the energy-momentum tensor.

Within this general context, we derive a simple relation between the
usual  coefficients $C_l$ of the multipole expansion
 of the CMB temperature fluctuations,
\beq
\left\langle{\delta T\over T}({\bf
n}){\delta T\over T}({\bf n}') \right\rangle_{{~}_{\!\!({\bf n\cdot
n}'=\cos\vartheta)}} ~~=~~~~
  {1\over 4\pi}\sum_\ell(2\ell+1)C_\ell P_\ell(\cos\vartheta)~,
\label{cor}
\eeq
and the fraction of critical density in the seeds. The relation
reads:
\beq
C_\ell = K ~~ \int_0^{\infty} d \log k ~~ |j_\ell (k \eta_0)|^2
 ~~ \Omega_{\sg}^2 (k, \eta_{re})~,
\label{final}
 \eeq
where
 $K$ is a  numerical  fudge factor $O(1)$, $\Omega_{\sg}(k,\eta)
\equiv \rho_{cr}^{-1} d \rho_{\sg} / d \log k$ is
the  seed   fraction of critical energy density per logarithmic
interval of frequency, evaluated at the conformal time $\eta$,
$\eta_0$ is the present (conformal) time and
 $\eta_{re}(k)$ indicates, for each comoving mode $k$, its  time of
re-entry, $\eta_{re} \sim k^{-1}$.  For the relevant
(large) scales, re-entry
occurs during the matter-dominated era. A crucial aspect of
(\ref{final}) is the appearance of $\Omega_\sg$ at a
$k$-dependent time (i.e. at re-entry), rather
than at a common (e.g. at recombination) time. This is because, in
the interesting
cases, the so-called ``integrated" Sachs-Wolfe (SW) contribution
\cite{SW} turns out to dominate over the ``ordinary" SW term.

In order to prove Eq.~(\ref{final}) we will proceed as follows.
We start
from a general formula expressing the  spectrum  of
primordial seed fluctuations in terms of the early, inflationary
evolution of the background. We then
 compute the inhomogeneities induced by this stochastic field in
the energy-momentum tensor as well as the Bardeen potentials.
Finally, we  estimate the large-scale temperature
anisotropy using the (total)
SW effect.  The result (\ref{final}) will thus
implicitly connect the observed CMB anisotropy to the very
early (possibly to the pre-big bang)
history of the universe.

This note is intended to give the essential
points in the argument and their main consequences.  For more
details on the calculation in a specific case we refer the reader to
our longer recent paper \cite{1}, where other kinds of seeds, as
well as the  case of massive
seeds, are also discussed. Further generalizations of
the massive case will be discussed in \cite{massive}.

Our computation of the fluctuations of $\si$ follows closely  the
general approach of \cite{BGV}. In a conformally flat metric the
effective action (\ref{action}) becomes:
\beq
S_{eff} =  {1 \over 2} \int d^3 x d \eta ~ S~\left[(  \si')^2 -
(\nabla \si)^2\right]~,
\label{action1}
\eeq
where  a prime stands for derivative with
respect to conformal
time $\eta$, and the so-called pump field $S$  is simply  $S\equiv
a^2 A$,   where  $a$ is the
scale factor of the homogeneous, isotropic, spatially-flat metric
resulting from
a long inflationary phase\footnote{It is important to note that
the results that we will obtain here are not
valid for conformally coupled fields, e.g for the electromagnetic
field, since these do not couple to the scale factor.}.
The corresponding effective Hamiltonian reads
\beq
H_{eff} = {1 \over 2} \int d^3 x ~\left[ S^{-1} \pi^2 + S (\nabla
\si)^2\right] ~~~,\label{ham}
\eeq
where $\pi = S \si'$
is the canonical variable conjugate to $\si$. The
Fourier modes of $\sigma$, when correctly normalized to
the vacuum before they ``exit" the horizon at the time
$|k\eta_{ex}| \sim 1$, are given by
\beq
\si (k, \eta) =  \frac{1}{\sqrt{k S}} ~~ e^{-i k\eta +
i\varphi_{{k}} }~~,~~~~~~ \pi (k, \eta)    = {\sqrt{kS}}
e^{-i k\eta +i\varphi'_{{k}} } ~~,  ~~~~\eta <   \eta_{ex} \sim -
k^{-1} ,
 \label{42}
\eeq
($\varphi_{{k}}, \varphi'_{{k}}$ are random phases, originating
from the random initial conditions). Furthermore, as far as
the computation of  energy
 spectra goes, fluctuations on superhorizon scales
 can be consistently truncated to their frozen
modes \cite{BGV} through
\beq
\si (k, \eta) =   \frac{1}{\sqrt{k S_{ex}}} ~~ e^{i\varphi_{{k}} }
~~,~~~~~   \pi (k, \eta)  =
{\sqrt{kS_{ex}}}  e^{i\varphi'_{{k}} } ~~,  ~~~~~\eta_{ex}
 < \eta <\eta_{re} \sim k^{-1} .
\label{43}
\eeq
The matching at re-entry finally gives, for $ k \eta>1$,
\begin{eqnarray}
\si (k, \eta)  &=&
\frac{1}{\sqrt{k S}} \left[\left(S_{re}\over S_{ex}\right)^{1/2}
\cos (k\eta)~ e^{i\varphi_{\vec{k}} }  +  \left(S_{ex}\over
S_{re}\right)^{1/2}
\sin(k\eta)~ e^{i\varphi'_{\vec{k}} }  \right]\, , \nonumber\\
\pi (k, \eta)  &=&
{\sqrt{kS}} \left[ \left(S_{ex}\over S_{re}\right)^{1/2}
\cos (k\eta)~ e^{i\varphi'_{\vec{k}} } - \left(S_{re}\over
S_{ex}\right)^{1/2}
\sin(k\eta)~ e^{i\varphi_{\vec{k}} }\right]~~.
\label{44}
\end{eqnarray}

 A nice feature of these results is their generality.
They hold for any kind of background and irrespectively of
whether a perturbation re-enters during the matter- or the
radiation-dominated epoch. Furthermore, these equations  respect an
invariance \cite {BGV}  of   cosmological perturbations
under the duality transformation $S \rightarrow S^{-1},
\nabla \si
\leftrightarrow \pi$.  For the sake
of simplicity, we shall consider here the case of a growing pump
field, keeping only the leading terms (those proportional to
$S_{re}/
S_{ex} \gg1$) in the fluctuations. This will result in
simpler formulae at the price of losing manifest  duality.

 The basic information  to be extracted from the preceding
formulae
 is the stochastic (spatial) average of $\si$:
\beq
\langle \si(\bk) \si^\ast(\bk')\rangle= {(2\pi)^3}
\da^3(k-k')\Sg(\bk, \eta)
\label{sstoch}
\eeq
where, according to Eqs. (\ref{43}) and (\ref{44}),
\bea
\Sg(\bk, \eta) = (k S_{ex})^{-1}  ~~~~,~~~~ k \eta <1 ~~,
\nonumber \\
\Sg(\bk, \eta) = (k S_{ex})^{-1} {S_{re}\over
S(\eta)}  ~~~~,~~~~ k \eta >1
~~. \label{stoch}
\eea
Equations  (\ref{sstoch}) and (\ref{stoch})
 allow us to compute the correlation functions of the
 seed energy-momentum tensor:
\beq
T_{\mu\nu}^{(\sigma)} \equiv T_{\mu\nu} = {S\over a^2}\left  
[\partial_{\mu} \si  \partial_{\nu}
\si - {1 \over 2} ~~g_{\mu\nu} (\partial_a \si)^2 \right].
\label{tmunu}
\eeq

Let us start with the average energy distribution
$d \rho_{\sg}(k) / d \log k= (k^3/a^4)\langle H \rangle$ which,
after   re-entry, can be computed from the Hamiltonian
(\ref{ham}) as \cite{BGV}:
\beq
{d \rho_{\si}(k) \over d \log k}
\simeq \left(k\over a \right)^4  {S_{re} \over
S_{ex}}(k)   \theta(k_1 -k).
\eeq
The end-point of the spectrum $k_1$ is the maximal amplified
frequency (the frequency that re-entered just after exiting), for
which just one quantum is produced per unit phase space. Above
$  k_1$ the
spectrum is exponentially depressed and we thus neglect it. Below
$k_1$, we can express $\rho_{\si}$ in  units of critical energy
density, $\r_c=3H^2/(8\pi G)$ as

\beq
\Omega_{\sg}(k,\eta)
 \simeq G\left({k^4\eta^2\over a^2}\right)
	\left({a_{re}\over a_{rad}}\right)^2\left( S_{rad}
\over S_{ex} \right)~
	\simeq G \left(k\over a_{re}\right)^2
\left( a_{re} \over a_{rad}   \right)^2\left( S_{rad}
\over S_{ex} \right)~ \left( a_{re} \over a \right)~.
\label{omega}
\eeq
We have denoted by $rad$ the beginning of the
radiation era, and
 we have assumed the background field $A$ to be  constant for
$\eta >   \eta_{rad}$.  Also, we
have limited our attention to scales relevant to the COBE DMR data,
which re-enter
during the matter-dominated era.  The
suppression   factor $(a_{eq} / a)$, naively expected for massless  
particles,
 is actually replaced by
$(a_{re} /  a )$. This is due to the additional amplification of modes
which are still outside  the horizon during (part of) the
matter-dominated phase.  In particular, just at   re-entry, we
find:
 \beq
\Omega_{\sg}(k,\eta_{re}) \simeq G \omega^2 \left( a_{re}
\over a_{rad}   \right)^2
\left( S_{rad}
\over S_{ex} \right)~.
\label{omegare}
\eeq
Clearly, some condition has to be imposed on the
behaviour of $S$ during inflation to ensure that $\Omega_\sg \ll 1$ at
all times.

Let us consider next the fluctuations of the various
components of the energy-momentum tensor and, in particular,
their power spectra $P_{\mu}^{\nu}$ defined by (no sum over
$\mu, \nu$ being implied): \beq
\langle T_\mu^\nu (x)  T_\mu^\nu (x')
\rangle- \langle T_\mu^\nu (x)\rangle \langle  T_\mu^\nu (x')
\rangle = \int {d^3 k \over (2\pi k)^3}e^{i{\bf k}\cdot ({\bf x-x'})}
P_{\mu}^{\nu}(k)~.
 \eeq
One easily finds that all the relevant components of
$P_{\mu}^{\nu}$
behave similarly,  and are controlled by
 a convolution of the form:
\beq
P_{\mu}^{\nu}(k,\eta) \sim  \left( S \over a^2 \right)^2~
\left(k^3\over a^4 \right)  \int d^3 p~ p^2
|k-p|^2 \Sigma(p) \Sigma(k-p) .
\eeq
Using Eqs. (\ref{stoch}), it is not hard to analyse the various
integration regions in $p$ in the above integral while always
keeping  $k \eta   \le 1$. In the region
$0< p < \eta^{-1}$ the integrand is proportional to $dp~ p^4~
S_{ex}^{-2}(p) $.
Imposing
that seeds  never be dominant makes this integrand peaked  at
its {\em upper} end.
On the other hand, in the region $\eta^{-1} <p <k_1$
the integrand behaves as $ dp~ p^{-4}~ ( d \rho_\sg / d \log
p)^2$. If the seed spectrum $\Om_\sg(p)$
grows with a small enough power of $p$, i.e. smaller
than $3/2$, this part of the integral is    dominated by its
{\em lower} end. In the opposite case it is dominated by the
cutoff   region $p \sim
k_1$ with uninteresting consequences \cite{1}.  In conclusion, in
the   interesting cases,
the integral  is dominated by the contribution around $p \sim
\eta^{-1}   \ge k$,  giving the
following white-noise spectrum for the energy density:
\beq (P_0^0)^{1/2} \sim {k^{3/2}\over a^{4} \eta^{5/2}} {S(\eta)
\over S_{ex}(p\sim \eta^{-1})} ~~\sim~~ (k/a)^4 ~~ (k
\eta)^{-5/2}~~ {S(\eta) } S^{-1} (-\eta)\;\; ,
\label{whitenoise}
\eeq
where positive and negative values of $\eta$ correspond,
respectively,  to the standard decelerated and
accelerated (inflationary) phases.

According to standard cosmological perturbation theory \cite{6a},
the spectrum of the Bardeen potentials $\Phi, \Psi$,
parametrizing the scalar fluctuations of
the metric in a gauge-invariant way, is related to
$P_0^0$ by\footnote {In general there can be additional
``compensation'' factors $(k \eta)^2$ appearing in this formula
\cite{1}. However, they will not matter since, in the end, we will
evaluate everything at $k\eta \sim1$.}
\beq
k^{3/2} |\Psi - \Phi|(k, \eta) \simeq G~ (a/k)^{2} (P_0^0)^{1/2}.
\label{BardP}
\eeq
Recalling  that $S\sim a^2$ for $\eta > \eta_{rad}$, we obtain
\bea
k^{3/2} |\Psi - \Phi|(k, \eta) &\simeq &  G
(k \eta)^{-5/2}\left(k\over a\right)^2{S(\eta) \over S_{rad}}~
{S_{rad} \over S(-\eta)}~
\simeq ~  G (k \eta)^{-5/2}\left(k\over a_{rad}\right)^2
{S_{rad}  \over S(-\eta)} \nonumber \\
 &\simeq & (H_1/M_P)^2~(k \eta)^{-5/2} (k/k_1)^2
\left[S_{rad}/S(-\eta)\right]~,
\label{Bardeen}
\eea
where $H_1 \equiv k_1/a_{rad}$ is the Hubble parameter at the
beginning of the
radiation era.
 Equations (\ref{Bardeen}) and (\ref{omega}) together
provide the interesting relation:
\beq
k^{3/2} |\Psi - \Phi|(k, \eta) \simeq  (k \eta)^{-5/2}
\left[S_{ex}(k)/S(-\eta)\right] \Omega_\sigma(k, \eta_{re})
\label{relation}
\eeq
and, in particular:
\beq
k^{3/2} |\Psi - \Phi|(k, \eta_{re}) \simeq \Omega_\sigma(k, \eta_{re})
\simeq (H_1/M_P)^2~(k/k_1)^2~\left[S_{rad}/S_{ex}(k)\right].
\label{relpart}
\eeq

At this point we insert the above result in the formula of the SW
effect, which is known to dominate the temperature
anisotropies at large angular scales, $\ell \ll 100$. Combining the
so-called ``ordinary'' and ``integrated'' SW contributions, a
standard analysis  \cite{1,Durrer} yields:
\beq C_\ell^{SW} =
{2\over\pi}\int {dk\over k}
\left\langle\left[\int_{k\eta_{dec}}^{k\eta_0}
k^{3/2} (\Psi   -\Phi)({\bf k},
\eta)j_{\ell}'\left(k\eta_0-k\eta\right) d(k \eta)\right]^2\right\rangle  ~,
\label{Cell}
 \eeq
where $j_l$ are the usual spherical Bessel functions
and $\eta_0, \eta_{dec} $ are, respectively, the present time and
the time of decoupling between matter and radiation (a prime stands here 
for the derivative of the Bessel function with respect to its
argument).
We exploit the previously determined $\eta$-dependence of the
Bardeen potentials, assuming that, after re-entry, these  potentials
are dominated by a cold dark matter (CDM) component and are therefore  
 constant.
We find \cite{1}  that the $\eta$ integral in Eq.~(\ref{Cell}) is  
dominated by
the region $k \eta \sim   1$, leading to:
\beq
C_\ell^{SW} \sim \int
d~(\log k) \left \langle \left[k^{3/2} (\Psi -\Phi)({\bf k},
\eta_{re})j_{\ell}\left(k\eta_0\right)\right]^2 \right\rangle  .
\label{Cell2}
 \eeq
Inserting (\ref{relpart})  we immediately recover the desired
result (\ref{final}). Note that
temperature fluctuations are controlled, for each scale $k$, by the
value of the
Bardeen potentials at the time it re-enters the horizon. Roughly:
\beq
(\Delta T/ T) (k) \sim (\Phi - \Psi) (\eta)|_{\eta \sim k^{-1}} \sim
\Omega_\sg (k, \eta_{re})~.
\eeq
In this way, the $\eta$ dependence of $(\Phi - \Psi)$ gets
translated into a $k$ (or $l$) dependence of the temperature
fluctuation spectrum.  Thus a scale-invariant $\Omega_\sg$ leads
to   scale-invariant Harrison-Zeldovich \cite{7a} spectrum of CMB
fluctuations.

For a simple power-law behaviour of the pump field,
$S_{rad}/S_{ex}(k) =  (k/k_1)^{\alpha-2}$, Eq.~(\ref{Cell2})
can be integrated analytically with the result:
\beq
 C_{\ell}^{SW} \approx K
(k_1\eta_0)^{-2\al}\left(H_1\over M_p\right)^4{\Ga(2-2\al)\over
4^{(1-\al)}\Ga(3/2-\al)} {\Ga(\ell+\al)\over\Ga(\ell + 2-\al)}.
~~~~~~~ \label{simple} \eeq
Comparing (\ref{simple}) with the standard inflationary result for CDM
\cite{JamesBond}, where the spectral index $n$ is defined
by \cite{JamesBond}
\beq
C_\ell^{SW} \propto {\Ga(\ell -1/2+ n/2)\over\Ga(\ell+5/2-n/2)} ~,
	\label{inflat}
\eeq
leads to the identification $(n-1)= 2\al$. More generally,
we can relate an effective (i.e. $k$-dependent) spectral index $n_{eff}$ 
to the behaviour of the pump
 field during inflation
via the relation:
\beq
(n_{eff}-1)/2 = \al_{eff} \equiv 2 - \left[d \log S_{ex}(k)/d \log  
k\right].
\label{indexrel}
\eeq
The nearly scale-invariant spectrum, measured by the DMR
experiment aboard the COBE satellite \cite{smootscott}, requires
\beq
 0.8\le n_{eff} \le 1.4 \label{259a}
\eeq
and thus, allowing for  generous error bars,  COBE's observations  imply
 \beq
 -0.1\le \al_{eff} \le 0.2
\label{alpourga0}
\eeq
in the very small $k$ region. Thus, through the definition of  
$\al_{eff}$ in
Eq. (\ref{indexrel}), one is able to relate COBE's data to the  
early-time evolution
of the pump field.

In this paper we have concentrated our attention on scalar perturbations. 
However, since the seeds
are of second order in the scalar field, we also expect the presence
of vector and tensor perturbations with roughly similar amplitudes.

Turning to the absolute normalization, we see from Eq.~(\ref{simple}) 
 that it  is controlled to a large extent  by the
crucial parameter $(H_1/M_P)^4$. The appearance of the fourth power of 
$H_1/M_P$ rather than the (more usual) second power is precisely
the   consequence of using  seeds --- rather than  first-order
fluctuations of the scalar field --- for
generating anisotropies. Thus $\Delta T/T$ goes like the square of
the original fluctuations. For the same reason, although the
fluctuations of $\si$   are expected to be
Gaussian, some non-Gaussianity is expected in the
fluctuations of $\Delta T/T$, since they are
sourced by the seed energy-momentum tensor, which is
quadratic in the Gaussian variable $\si$. Thus, we rather expect
$\Delta T/T$ to obey a $\chi^2$ statistics (note that this is one of
the few non-Gaussian examples where we really have a handle on the
statistics).

Let us finally turn our attention to a specific example of our new
mechanism, that of
the universal axion in pre-big bang (PBB) cosmology. The universal
axion of superstring theory is just the (pseudo-scalar)
partner of the dilaton in the string effective action, and is
massless in perturbation theory because of a Peccei-Quinn (PQ)
symmetry. While the dilaton is expected to acquire a
mass as soon as supersymmetry is broken, the axion could remain
massless, or almost   massless, because of its Nambu-Goldstone
origin. Although the PQ symmetry is broken by
instantonic effects, in the presence of various axions coupled
to the same topological current, a  linear combination, mainly lying
along the invisible axion's axis, is expected to remain very light or  
massless
 (for the purpose
of this work a mass of order $H_0$ can be considered
to be zero). Such a light particle, being a gravitationally  coupled  
pseudo-scalar,
should not lead to phenomenological difficulties.

The  field $A$ of Eq.~(\ref{action}) turns out to be, in this case,
$e^{\phi}$ (where
$\phi$ denotes the dilaton) and is related to the effective Newtonian
constant,  in the conventions used in PBB cosmology
\cite{PBB}, by $G_N^{eff} \sim e^{\phi}$. The pump field $S$ is
thus   $a^2 e^{\phi}$, and grows very fast during PBB inflation
since both $a$ and $e^{\phi}$   have accelerated
behaviour (on the contrary, the pump field for dilaton
and gravity-wave perturbations is $a^2 e^{-\phi}$, and the two
factors tend to   cancel out,  giving
too little power at large scales \cite{5b}). In the axion case the
exponent $\al$ of Eq.~(\ref{simple}) can be evaluated \cite{Copeland}  
from the
known background   solutions \cite{PBB}. As
noticed in \cite{Buon}, the desired value $\al=0$ is reached, in
particular, for a
highly symmetric, ten-dimensional PBB background in which the
six   extra dimensions
evolve like the three ordinary ones (up to an irrelevant $T$-duality  
transformation). Precisely
in this case, a scale-invariant spectrum for $\Delta T/T$ will result.

Concerning the overall normalization,  controlled by
$(H_1/M_P)^2$, we note that
in the PBB scenario the inflationary scale $H_1$ is typically of
order of the string scale $M_s$,   usually taken to be
around $5\times 10^{17}$ GeV. Typically,  $(H_1/M_P)^2$
thus  varies
between $10^{-2}$ and $10^{-4}$. We   have to add the
fudge factor $K$, which is hard to evaluate  precisely, but is
expected to contain factors  like  $(16 \pi ^2)^{-1}$. Thus,
amusingly enough, the right order of magnitude \cite{33}  for $C_2$  
($C_2 \sim
10^{-10}$) may come out naturally from $K (M_s/M_P)^4$ (taking
also into account the possibility that the  spectrum  be slightly
tilted, see \cite{1} for a quantitative discussion).

In conclusion, irrespectively of its possible model-dependent
origin, we believe that a cosmic background of massless pseudo-scalar
fluctuations may provide a consistent and interesting
explanation of the anisotropies observed in the CMB
temperature, at large angular scales. It is unclear,
at present, whether such an axion-induced anisotropy may lead
to significant differences  in
the acoustic peak structure of the
CMB anisotropy spectrum at smaller angular scales. If it does, this  
(plus possibly
some non-Gaussianity of the fluctuations) should
allow tests of our axionic-seed mechanism  through the high-precision
 measurements planned for the near future \cite{35}.
The discussion of this possibility is postponed to further work.

\end{document}